\documentclass{elsart}
\usepackage[dvips]{graphics}
\begin{document}
\bibliographystyle{unsrt}

\journal{Astroparticle Physics}

%%%%%%%%%%%%%%%%%%%%%%%%%%%%%%%%%%%%%%%%%%%%%%%%%%%%%%%%%%%%%%%%%%%%%%%%%%%%%%%
% Joe's special commands and standards
\hyphenation{Cher-en-kov}
\newcommand{\unit}[1]{\ensuremath{\,\mathrm{#1}}}
\newcommand{\gcm}[1]{\ensuremath{#1\unit{g\,cm}^{-2}}}
\renewcommand{\etal}{\emph{et al.}}
\newcommand{\Xmax}{\ensuremath{X_{max}}}
\newcommand{\meanXmax}{\ensuremath{\langle X_{max} \rangle}}
\newcommand{\lnA}{\ensuremath{\ln(A)}}
\newcommand{\meanlnA}{\ensuremath{\langle \ln(A) \rangle}}
\newcommand{\corecut}{30\,m}
\newcommand{\zencut}{\ensuremath{9^\circ}}
\newcommand{\exposure}{\ensuremath{1.83\times10^{10} \unit{m}^2 \unit{sr} \unit{s}}}

% Use \blancafig to make a figure, left-right centered on page
\newcommand{\blancafig}[2]{
  \begin{center}
    \resizebox{#2}{!}{\includegraphics{#1}}
  \end{center}
}

% Use \bigfig to make a full-width figure, left-right centered on page
\newcommand{\bigfig}[1]{\blancafig{#1}{\textwidth}}

%%%%%%%%%%%%%%%%%%%%%%%%%%%%%%%%%%%%%%%%%%%%%%%%%%%%%%%%%%%%%%%%%%%%%%%%%%%%%%%
\begin{frontmatter}
\title{A Measurement of the Cosmic Ray Spectrum and Composition at the Knee}

\author[EFI]{J.W. Fowler\thanksref{Corres}}
\author[EFI,Adler]{L.F. Fortson}
\author[Utah]{C.C.H. Jui}
\author[Utah]{D.B. Kieda}
\author[EFI]{R.A. Ong}
\author[EFI]{C.L. Pryke}
\author[Utah]{P. Sommers}

\address[EFI]{The Enrico Fermi Institute, University of Chicago, 
              5640 Ellis Avenue, Chicago, Illinois 60637-1433, USA}
\address[Adler]{Dept.\ of Astronomy, Adler Planetarium and Astronomy
Museum, Chicago, Illinois 60605, USA}
\address[Utah]{High Energy Astrophysics Institute, Dept.\ of Physics,
University of Utah, Salt Lake City, Utah 84112, USA}
\thanks[Corres] {Corresponding author: Dept.\ of
              Physics, Princeton University, Princeton, New Jersey
              08544, USA.  Email: jfowler@princeton.edu.
%	      Phone: 1-609-258-4365.
	}

%%%%%%%%%%%%%%%%%%%%%%%%%%%%%%%%%%%%%%%%%%%%%%%%%%%%%%%%%%%%%%%%%%%%%%%%%%%%%%%
\begin{abstract}
The energy spectrum and primary composition of cosmic rays with energy
between $3\times 10^{14}$ and $3\times10^{16}\unit{eV}$ have been
studied using the CASA-BLANCA detector.  CASA consisted of
957 surface scintillation stations; BLANCA consisted  of 144
angle-integrating Cherenkov light detectors located at the
same site.  CASA measured the charged particle distribution of air
showers, while BLANCA measured the lateral distribution of Cherenkov light.
The data are interpreted using the predictions of the CORSIKA air shower
simulation coupled with four different hadronic interaction codes. 

The differential flux of cosmic rays measured by BLANCA exhibits a
knee in the range of 2--3\,PeV with a width of approximately 0.5
decades in primary energy.  The power law indices of the differential
flux below and above the knee are $-2.72\pm0.02$ and $ -2.95\pm0.02$,
respectively.

We present our data both as a mean depth of shower maximum and
as a mean nuclear mass.
A multi-component fit using four elemental species suggests the
same composition trends exhibited by the mean quantities, and also
indicates that QGSJET and VENUS are the preferred hadronic interaction
models.  We find that an initially mixed composition turns lighter
between 1 and 3\,PeV, and 
then becomes heavier with increasing energy above 3\,PeV\@.  

\end{abstract}

\end{frontmatter}

%%%%%%%%%%%%%%%%%%%%%%%%%%%%%%%%%%%%%%%%%%%%%%%%%%%%%%%%%%%%%%%%%%%%%%%%%%%%%%%
\section{Introduction} \label{sec.intro}

The all-particle energy spectrum of cosmic rays can be described by
a steeply falling power law over many decades of energy. The smoothness of
this drop in intensity with energy is broken by a change in index of the 
power law just above $10^{15}\unit{eV}$. While the origin of this break 
(referred to as the ``knee'') is not yet fully understood, the
prevailing theoretical models describe the knee as a result of the
energy limit for particle acceleration in supernova
shocks~\cite{Cesarsky,Drury94}. Further, these models predict that
the composition of the primary cosmic rays should change from proton 
(or ``light''-nuclei) dominated to iron (or ``heavy''-nuclei) dominated 
as energy increases through the region of the knee.  Measuring a
composition trend from light to heavy would lend support to
the supernova shock acceleration picture. 

%\P: what makes this so hard to measure
Determining the composition of cosmic rays with energies greater than 
$10^{15}\unit{eV}$ is a notoriously difficult problem. To detect
primary cosmic rays above this energy directly by satellite
experiments requires an unacceptable launch payload volume. Similarly, 
stratospheric balloon-borne experiments are limited by volume and flight 
time in their collection of primary particles.
Thus, to investigate the composition of cosmic rays at the knee, we 
must rely on ground-based detection of air showers
generated  by the primary cosmic rays.

%\P: The BLANCA concept (LDF->E&Comp)
The Cherenkov light emission from the charged particle component of 
an air shower provides an integrated measurement of the longitudinal 
development~\cite{Brennan58,Chudakov60}.  One approach is to sample
the Cherenkov lateral distribution, the photon density
as a function of distance from the air shower core.
The Cherenkov intensity is proportional to the primary energy,
while the slope of the lateral distribution is related to the depth
of maximum shower development --- and hence to the mass of the primary
cosmic ray nucleus.
Therefore measuring a large number of Cherenkov lateral distributions can
provide information on how the composition changes with
energy~\cite{patterson_hillas}.
Previous attempts to exploit this fact at the knee
include~\cite{Dawson89,airobicc,vulcan}.
% Who was the first to
% suggest using the C lat dist to get composition information?
% (As opposed to just measuring the energy spectrum.) Do the Brennan
% and Chudakov papers suggest it?

% Why build a Cherenkov array?
Optical photons suffer little absorption as they travel
through the atmosphere.
This means that the Cherenkov lateral distribution is much
broader than that of charged particles.
Additionally their numerical density is much higher.
Thus it is possible to make high signal-to-noise
measurements of Cherenkov lateral distributions using an array
of detectors with smaller area and wider spacing than would
be required for equivalent measurements of charged particles.

%\P: Where BLANCA fits in, what its aims are
To obtain high quality Cherenkov lateral distribution data the
Broad Lateral Non-imaging
Cherenkov Array (BLANCA) was built at the Chicago Air Shower Array (CASA)
installation in Dugway, Utah. Using CASA as the cosmic ray trigger, BLANCA
operated on clear, moonless nights in 1997 and 1998.
In the following analysis we use CASA to find the shower core position and
arrival direction and BLANCA to make a precision measurement
of the Cherenkov lateral distribution.
This paper details the results obtained through these 
measurements on the energy spectrum and composition of cosmic rays in
the energy range between $3\times 10^{14}$ and $3\times10^{16}\unit{eV}$.  

% The dates in the previous paragraph are repeated in the next but one section

%%%%%%%%%%%%%%%%%%%%%%%%%%%%%%%%%%%%%%%%%%%%%%%%%%%%%%%%%%%%%%%%%%%%%%%%%%%%%%%
\section{The CASA-BLANCA Instrument} \label{sec.blanca}

\begin{figure}
\bigfig{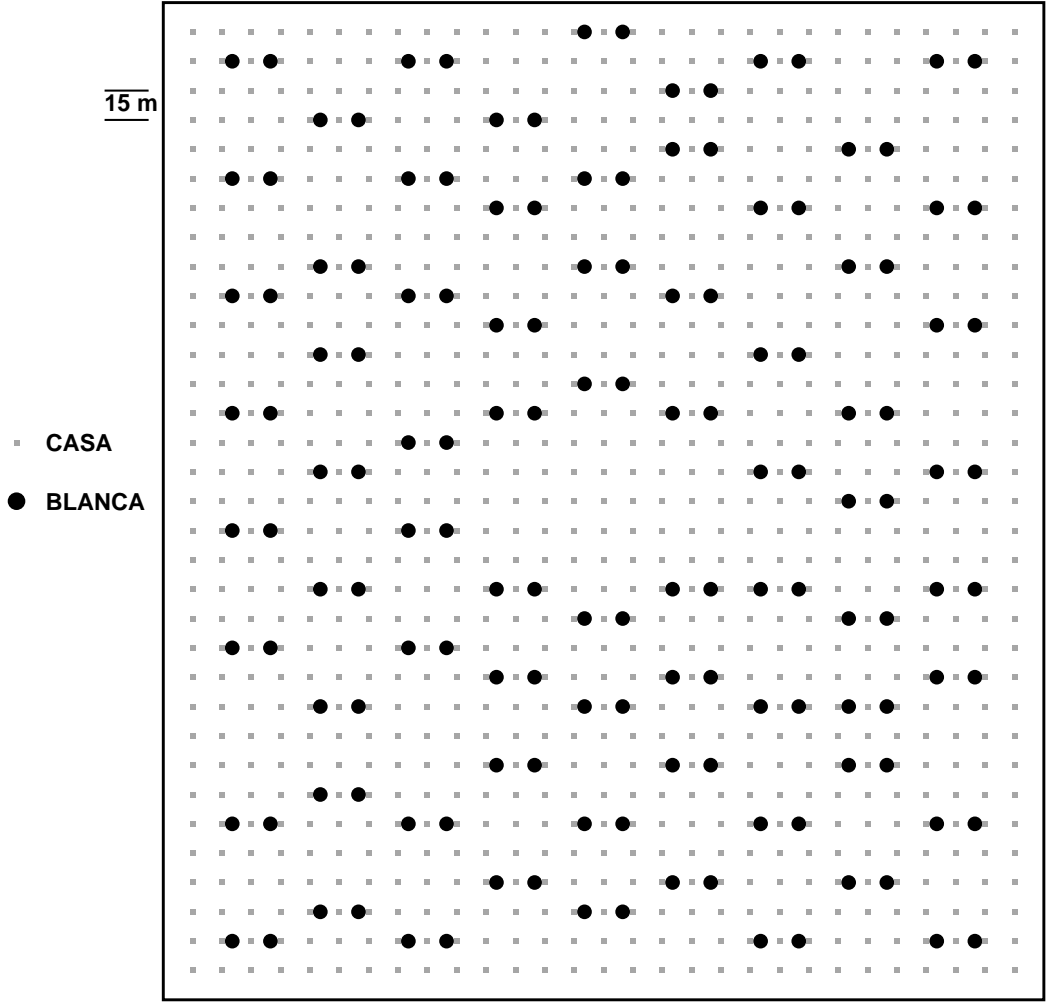}
\caption
{The CASA-BLANCA arrays.  The CASA scintillator stations were
placed at 15\,m intervals.  The 144 BLANCA air Cherenkov detectors filled the
array with characteristic separations of 35--40\,m.}
\label{fig.site_map}
\end{figure}

%\P: CASA and BLANCA site
The CASA-BLANCA instrument was located at the Dugway Proving Ground
near Salt Lake City, Utah, USA, under a mean atmospheric overburden of
$870\unit{g\,cm}^{-2}$.  During BLANCA runs,
CASA~\cite{casa_nim} consisted of 957 
scintillation counters which detected the charged particles in an air
shower.  The surface array covered approximately 0.2\,km$^2$. 
BLANCA~\cite{blanca_instrument2} consisted of 144
angle-integrating detectors which recorded the lateral
distribution of air shower Cherenkov light.  The BLANCA detectors were
not uniformly spaced but had an average separation of 35--40\,m.
MIA, an array of buried muon detectors at the same site, was not used
in this analysis.  Figure~\ref{fig.site_map} shows the site plan.

%\P: BLANCA detectors, optical and electronic
Each BLANCA detector contained a large Winston cone~\cite{winston}
which concentrated the light striking an 880\,cm$^2$ entrance aperture
onto a photomultiplier tube.  The 
concentrator had a nominal half-angle of 12.5$^\circ$ and truncated
length of 60\,cm.  The geometrical concentration ratio of this design
was 19, but losses in the system reduced the effective concentration
ratio to 15.  Lab measurements and simulations show that the
effective half-angle was actually $\sim 10^\circ$ because of a 6\,mm
gap between the photomultiplier and the cone.  The Winston cones were aligned
vertically with $\sim 0.5^\circ$ accuracy.  A two-output preamplifier
increased the dynamic range of the detector.  A typical BLANCA
unit had a detection threshold of approximately one blue photon per
cm$^{2}$. 

%\P: use of CASA for trigger
The Cherenkov array did not have a separate trigger system,
relying instead on triggers from CASA.  Cherenkov data were recorded for
all stations which exceeded a fixed threshold in coincidence with a surface
array trigger.  For showers in the BLANCA field of view, the CASA
trigger threshold imposes a minimum energy of $\sim 100\unit{TeV}$
on the Cherenkov array. However, to eliminate composition bias inherent
in the CASA trigger threshold,  we use only showers with an
energy of at least 200\,TeV as determined by BLANCA.

%%%%%%%%%%%%%%%%%%%%%%%%%%%%%%%%%%%%%%%%%%%%%%%%%%%%%%%%%%%%%%%%%%%%%%%%%%%%%%%
\section{Data Collection and Calibration} \label{sec.data}

%\P: Operating period, data cuts, exposure
CASA-BLANCA operated on 90 moonless nights between January,
1997 and May, 1998.  After removing periods of hazy or cloudy weather,
approximately 460 hours of Cherenkov observations remain.  Data cuts
require events to have at least five good Cherenkov measurements from
BLANCA and a reconstructed primary direction within \zencut\ of
zenith.  Events are also cut if the core location reconstructed by
CASA is outside the array or within 30\,m of the edge, because core
location uncertainty increases towards the edge.  The geometrical and
temporal cuts result in an exposure to
cosmic rays of \exposure.

%\P: Intercalibration
For each night of data the BLANCA detectors are intercalibrated using the
cosmic ray data itself to find their relative sensitivity to
Cherenkov light.  The intercalibration method depends on the circular 
symmetry of the Cherenkov light pool about the shower axis.   Two
detectors equidistant from a shower core receive, on average, equal  
Cherenkov photon densities, and any discrepancies can be attributed
to different detector sensitivities.  By averaging over suitable events
in a run, we find the sensitivity ratio of each possible pair
of BLANCA detectors.  A maximum-likelihood method is then used to
determine the set of 143 relative gains that best reproduces the
pairwise ratios~\cite{thesis}.  The reliability of this method was
verified with an \emph{in situ}\ intercalibration using a portable
stable blue LED flasher ($\lambda\sim 430\unit{nm}$).  The relative
detector sensitivities are distributed log-normally with typical RMS
of 0.4 in natural log.  The relative sensitivies are stable over
time except for occasional changes in detector hardware.

%\P: Absolute light calibration and time monitoring
The BLANCA absolute calibration also used the blue LED system as a
reference source.  The LED flasher was calibrated by using it to
produce single photoelectrons in a photomultiplier with good charge
resolution.  Two BLANCA detectors were then calibrated in a dark 
box using the reference LED.  The absolute calibration has  a 20\%
systematic uncertainty,  resulting mainly from the uncertain quantum
efficiency of the reference photomultiplier.  This absolute
calibration error produces a similar uncertainty in the overall
energy scale.  The spectral response of the BLANCA
photomultipliers was measured by the manufacturer for four
representative tubes.  As the relative response was similar for these
four detectors, it was also assumed to be the same for all BLANCA
detectors. 

%%%%%%%%%%%%%%%%%%%%%%%%%%%%%%%%%%%%%%%%%%%%%%%%%%%%%%%%%%%%%%%%%%%%%%%%%%%%%%%
\section{Air Shower Simulations} \label{sec.simulation}

%\P: CORSIKA and use of 4 models
The CASA-BLANCA analysis compares the Cherenkov measurements with air
showers simulated by the CORSIKA Monte Carlo version
5.621~\cite{corsika}.  We used 
the EGS4 and GHEISHA codes for the electromagnetic and low-energy
nuclear interactions.
Nucleus-nucleus interactions at air shower energies are well beyond
the reach of accelerator experiments.
Therefore we are forced to rely on hadronic interaction models
which attempt to extrapolate from the available data using different
mixtures of theory and phenomenology.
Several groups produce such models --- in this paper we have used four:
QGSJET~\cite{qgsjet}, VENUS~\cite{venus},
SIBYLL~\cite{sibyll}, and HDPM~\cite{corsika}. BLANCA data are
interpreted according to each model, indicating the
systematic errors that depend on the choice of interaction model.

%\P: Libraries and thinning
Simulated showers were produced for proton, helium, nitrogen, and iron
primaries in equal numbers.  Nitrogen was chosen to represent the
entire CNO group.  The shower libraries consist of 10,000 showers per
primary species, per hadronic model.
Primary energies are  distributed uniformly in 
$\log(E)$ between $10^{14}$ and $10^{16.5} \unit{eV}$\@.  Shower
directions are uniform in solid angle to a maximum zenith angle of
$12^\circ$.  To produce such large simulated air shower libraries, it
was necessary to employ the CORSIKA thinning option, which tracks
only a representative sample of shower particles below a threshold
energy.  In all simulations, this threshold was $10^{-4}$ times the
primary energy.  Studies of simulated 1\,PeV showers showed that at
this thinning level, the distributions of gross properties such as
the depth of shower maximum and Cherenkov slope and intensity 
were indistinguishable from those of unthinned showers~\cite{icrc_corsika}.

%\P: Scattering model, effects of scattering
Shower development and Cherenkov emission were simulated in Monte
Carlo assuming the U. S. standard atmosphere.  The CORSIKA
program was modified to include a complete model of atmospheric
scattering, both Rayleigh (molecular) and Mie (aerosol).  Although the
modified CORSIKA tracks scattered photons, few reach the ground
within the BLANCA field of view, effectively making scattering an
absorption process for Cherenkov photons.  The scattering losses
were generally similar in magnitude to the expected measurement errors
and were correlated with the depth of shower maximum.
Therefore, atmospheric scattering
must not be ignored in analyzing the BLANCA data.
On average, scattering reduces the Cherenkov intensity by
$\sim 20\%$ and increases the inner slope by $\sim 7\%$.
Scattering effects are smallest for late developing showers.  
Possible molecular absorption by oxygen and ozone was determined to
be small compared with the systematic error in the absolute detector
calibration and was consequently ignored.

%\P: Full detector simulation
The CORSIKA air showers were processed by a full BLANCA detector
simulation which includes the measured wavelength dependence and
angular response of the BLANCA detectors.  Other simulated effects
include detector alignment; unequal detector gains and saturation
levels; night sky background light; photomultiplier response; and
errors in the CASA core location and shower direction.  The detector
simulation produces ``fake data'' which is calibrated and fit
like the real data.  We have used this fake data to find the
optimum transfer functions for converting Cherenkov measurements to
air shower and primary cosmic ray parameters throughout this paper.

%%%%%%%%%%%%%%%%%%%%%%%%%%%%%%%%%%%%%%%%%%%%%%%%%%%%%%%%%%%%%%%%%%%%%%%%%%%%%%%
\section{The Cosmic Ray Energy Spectrum} \label{sec.spectrum}

%\P: Cherenkov LDF fits
For each air shower event, raw BLANCA data are converted to photon
densities, producing a Cherenkov lateral distribution.  We fit this
lateral distribution with an empirically motivated function which
matches both the real and simulated data.  The function is exponential
in the range 30\,m--120\,m from the shower core and a power law from
120\,m--350\,m.  It has three
parameters: a normalization $C_{120}$, the exponential ``inner slope''
$s$, and the power law index $\beta$:

\begin{equation}
C(r) = \left\{
   \begin{array}{ll}
   C_{120}\   e^{s(120\unit{m}-r)},     & 30\unit{m}<r\leq 120\unit{m} \\
   C_{120}\   (r/120\unit{m})^{-\beta}, & 120\unit{m}<r\leq
   350\unit{m} \\
   \end{array}  \right.
\label{eq.ldf}
\end{equation}

%\P: Energy transfer function method
The energy of each air shower is derived using only the $C_{120}$ and
$s$ parameters of the Cherenkov lateral distribution fit.  The outer
slope $\beta$ is not used, both because it correlates strongly with
$s$ and because it is subject to larger measurement errors.
The Monte Carlo fake data libraries (including detector 
simulation) are used to determine the relationship between measured
quantities and energy.  The energy depends primarily on
$C_{120}$, the Cherenkov intensity 120\,m from the core.  We fit the
logarithm of the energy as a quadratic function of $\log C_{120}$ (the
curvature is small, typically 0.005 decades$^{-1}$).  In all hadronic
models, $C_{120}$ grows approximately as $E^{1.07}$, because the
fraction of primary energy directed into the electromagnetic component
of the cascade increases with energy.

% Slope correction
According to the Monte Carlo, the quadratic function used to estimate
the energy works well for most
showers but has a bias such that energies are slightly underestimated
for showers with unusually large or small depths of maximum development.
Therefore, a small correction is applied to the energy estimate.  The
correction depends on the Cherenkov slope $s$ and on the shower zenith
angle.  The magnitude of this energy correction is less than 10\% for
85\% of BLANCA showers.

% Energy resolution
Energies derived from the data in this manner have an error
distribution which depends on the primary
mass and energy.  In general, the random errors on
reconstructing a single shower's energy are comparable to the
systematic uncertainty due to the unknown composition.  Assuming a
mixed cosmic ray composition, the BLANCA energy resolution for a
single air shower is approximately 12\% for a 200\,TeV shower, falling
to 8\% for energies above 5\,PeV.

% Spectrum figure
\begin{figure}
\bigfig{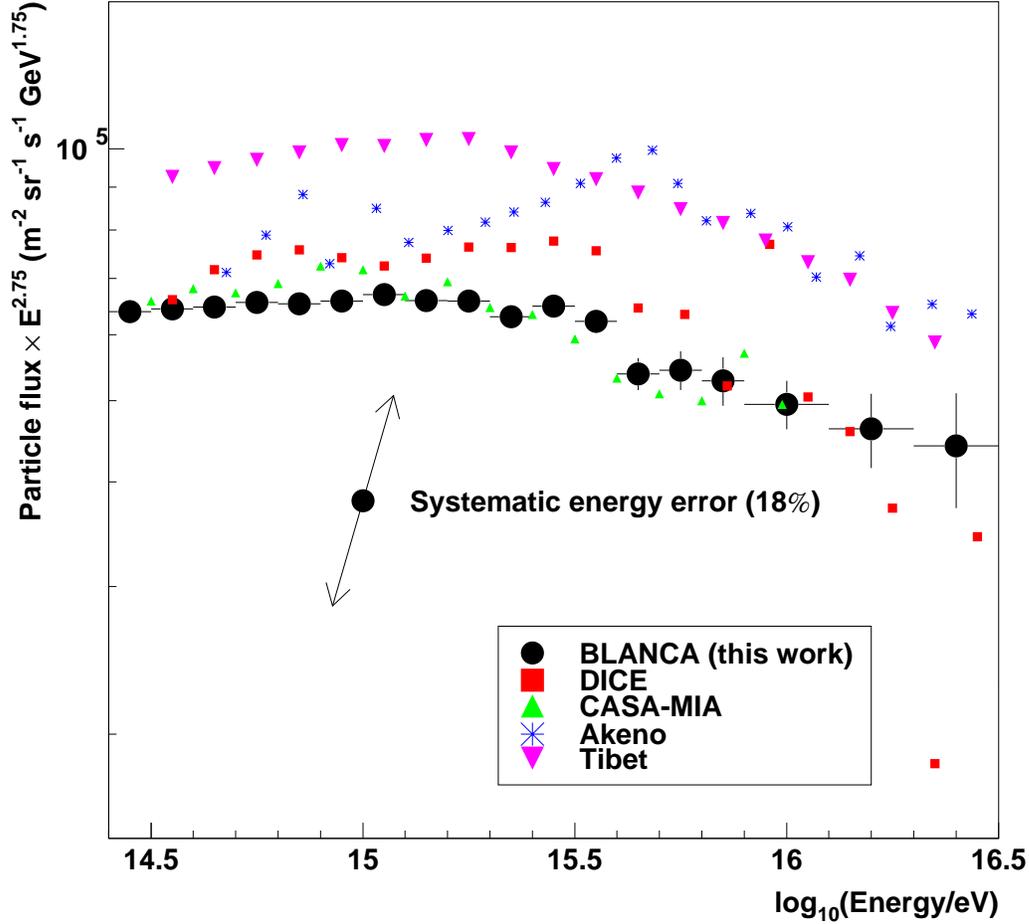}
\caption{The differential all-particle cosmic ray flux measured by
CASA-BLANCA (QGSJET energy).  Each data point represents the differential flux
scaled up by a factor of $(E/1\unit{GeV})^{2.75}$; the error bars represent
statistical (Poisson) errors only.  The diagonal
bar shows the effect of a possible systematic energy shift of $\pm18\%$.
%The fluxes have power law indices of $-2.72\pm0.02$ below 1\,PeV and
%$-2.95\pm0.02$ above 3\,PeV.  
Also plotted are the cosmic ray fluxes
reported by DICE~\cite{dice}, CASA-MIA~\cite{ande_spec},
Akeno~\cite{akeno} and Tibet AS$\gamma$~\cite{tibet}.}
\label{fig.spectrum}
\end{figure}

%\P: Discussion of spectrum plot and table
%\P: Make the point that this gives confidence in our calibration, etc
The differential all-particle cosmic ray flux measured by CASA-BLANCA
is shown in Figure~\ref{fig.spectrum}, scaled up by a factor of
$(E/1\unit{GeV})^{2.75}$ to emphasize the structure. 
Table~\ref{tab.spectrum} lists the values of the observed spectrum.
The energy spectrum is compared with that reported by several other groups.
Although the CASA-BLANCA, DICE, and CASA-MIA experiments shared some
instrumentation, their data sets and energy analysis methods are
entirely independent.  These three experiments show very good agreement
in their spectrum determination.  Most other results are consistent with
CASA-BLANCA, given the 20\% or larger energy systematic error typical of
air shower measurements.

% Knee fit functions and choice of $w$
\begin{figure}
\begin{center}
\resizebox{.7\textwidth}{!}{\includegraphics{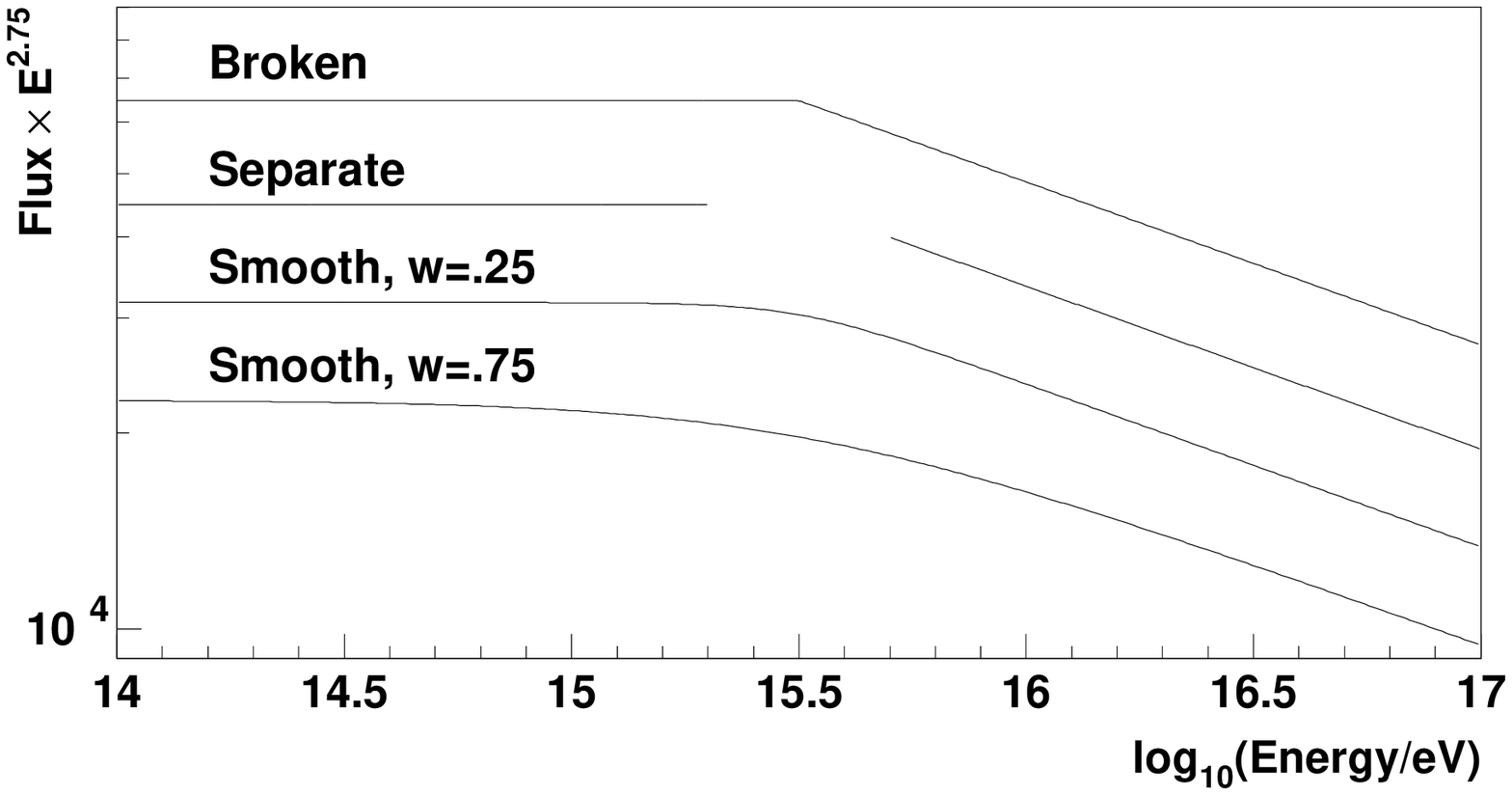}}
\resizebox{.7\textwidth}{!}{\includegraphics{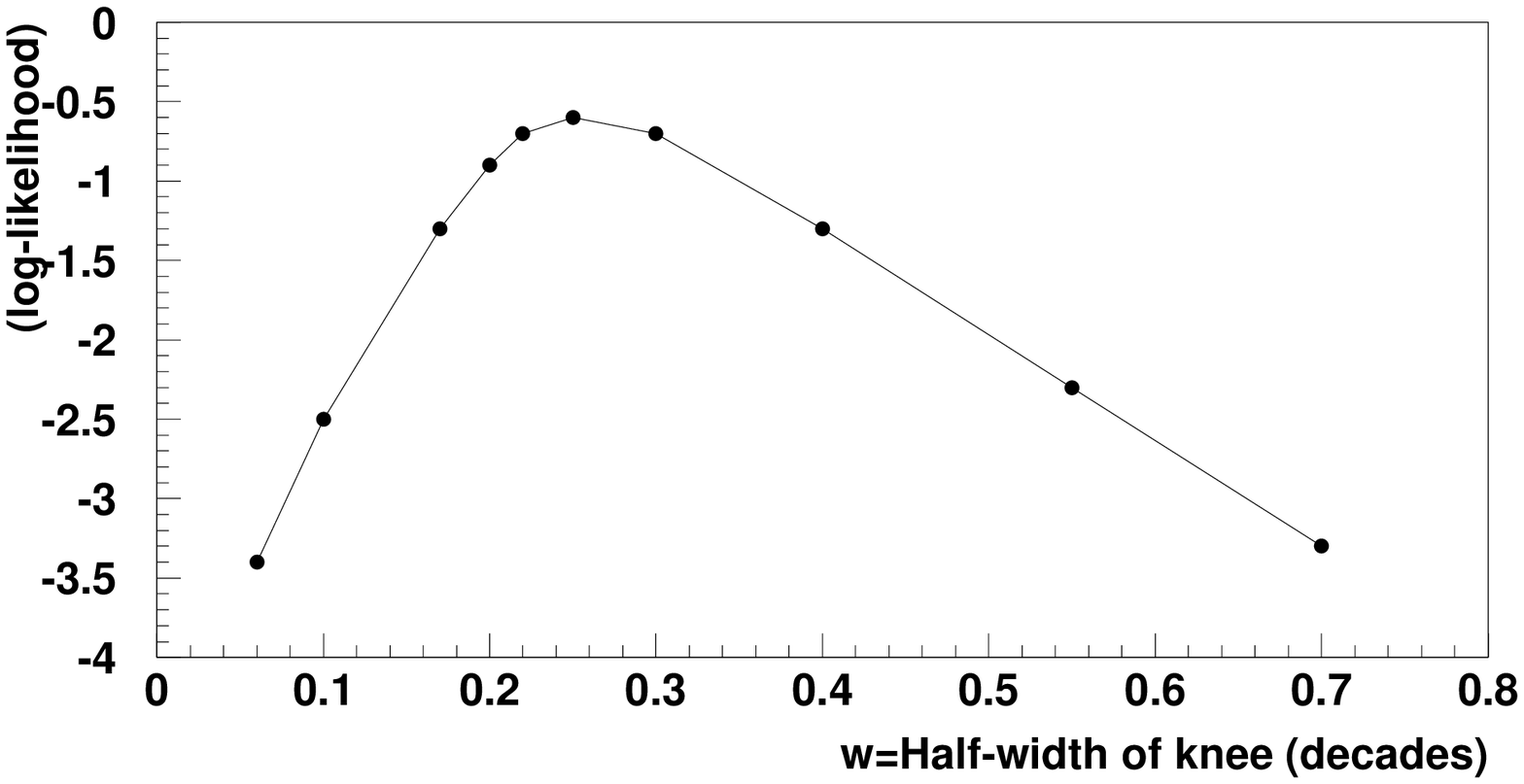}}
\end{center}
\caption{\emph{Top:} Several trial functions used to fit the cosmic
ray flux.  \emph{Bottom:} The log-likelihood for the smooth function at
a range of widths $w$.  The CASA-BLANCA data fit best to a knee with
$w=0.25$, or a full width of one-half decade.}
\label{fig.knee}
\end{figure}

% Discussion of QGS vs everything else
The spectrum shown in Figure~\ref{fig.spectrum} uses event energies derived
from the Cherenkov predictions of CORSIKA with the QGSJET hadronic
interaction model.  The data can also be interpreted using the other
available interaction models.  The alternate energy estimates lead to
spectra with no qualitatively different features.  Instead, they
amount only to a shift in energy scale of order 10\%, less than
the BLANCA instrumental energy scale uncertainty.  HDPM and VENUS
predict less Cherenkov light and hence assign higher energies than QGSJET
does, while the SIBYLL simulations lead to lower assigned energies.

%\P: Discussion of knee: position and index fits
Figure~\ref{fig.spectrum} contains the knee region of the
spectrum, near 3\,PeV.  The CASA-BLANCA spectrum exhibits a smooth
change rather than a sharp break here.  However, measurements over a wider
energy range show that the form of the cosmic ray spectrum is a power
law well above and below 3\,PeV.  Historically, many groups have found
the knee to be quite sharp.  In the spirit of the usual discussion of
the knee, we have fit several similar functions to the data 
(Figure~\ref{fig.knee}, top).  The fits find simultaneously the
position (energy) of the knee and the power law indices above and
below the knee.  A log-likelihood fit is performed in order to
account for the Poisson statistics of discrete events in a binned
energy distribution.  

\begin{figure}
\blancafig{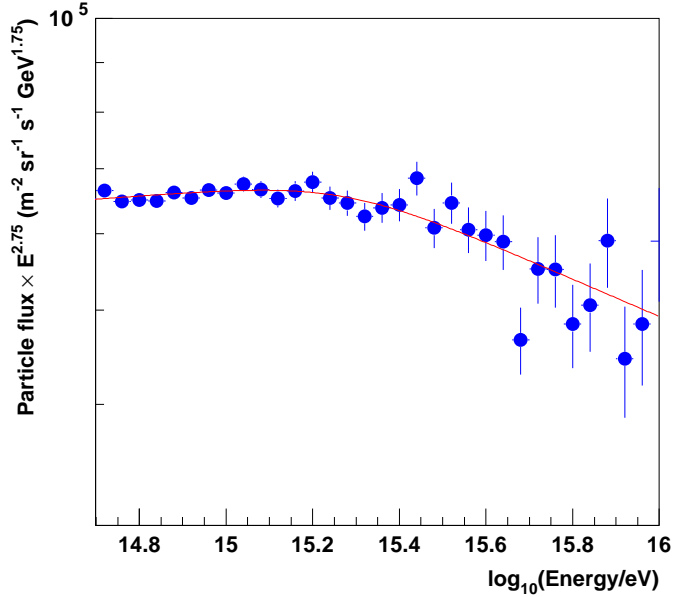}{.7\textwidth}
\caption{The CASA-BLANCA energy spectrum shown with more detail in the knee
region.}
\label{fig.knee_zoom}
\end{figure}

% Smooth knee function
Of the trial functions, the smoothly changing power law fits the
BLANCA data best: 
\begin{equation}
J(E) = J_k \left(\frac{E}{E_k}\right)^\alpha 
\left[1+\left(\frac{E}{E_k}\right)^\frac{1}{w}\right]^{(\beta-\alpha)w}
\label{eq.knee}
\end{equation}
$E_k$ is the energy at the center of the transition, \emph{i.e.} the
knee energy.  For $E \ll E_k$, the function is a power law with index
$\alpha$, while the spectral index becomes $\beta$ for $E \gg E_k$.
Parameter $J_k$ sets the normalization at the knee.  The fifth
parameter, $w$, is the half-width in decades of the transition region.  
The lower panel of Figure~\ref{fig.knee} shows how the log-likelihood
depends on the choice of $w$.  The data favor a knee
one-half decade wide ($w=0.25$).  The best-fit knee energy is
$2.0^{+0.4}_{-0.2}\unit{PeV}$, with power law indices of
$\alpha=-2.72\pm0.02$ and $\beta=-2.95\pm0.02$.
Figure~\ref{fig.knee_zoom} shows the energy spectrum near the knee
using fine bins 0.04 decades wide as well as the best fit curve.

%%%%%%%%%%%%%%%%%%%%%%%%%%%%%%%%%%%%%%%%%%%%%%%%%%%%%%%%%%%%%%%%%%%%%%%%%%%%%%%
\section{Depth of Shower Maximum} \label{sec.xmax}

%\P: Why we are quoting results for a quantity which is not
% of any intrinsic astrophysical interest.
The depth of shower maximum (\Xmax) is an important characteristic
of air shower development which, for given energy, is related to the mass
of the primary particle.
However, like all air shower parameters, the relationship
depends on the choice of uncertain high energy hadronic interaction
model parameters.
Imaging experiments such as Fly's Eye~\cite{flys_eye}, and to a lesser
extent DICE~\cite{dice}, measure \Xmax\ rather directly.
The importance of \Xmax\ for various other types of ground-based
experiment has long been known~\cite{PattersonHillas}; measured
parameters can often be translated into the depth of shower
maximum in a way which is rather independent of the hadronic model.
Therefore \Xmax\ provides a useful middle ground on which experiments
may publish and compare their results.

% This bit's not necessary except to help an outsider understand
% the mean Xmax figure
The mean \Xmax\  for a given primary type
grows logarithmically with energy at an approximate elongation
rate of \gcm{80} per decade, although this value depends on
the hadronic model used.  The expected  \Xmax\   is similar for
two primaries of different mass if they have equal energy per nucleon.

% Fit procedure similar to energy simplefit, libraries
To determine the optimum transfer function for converting Cherenkov
lateral distributions 
into \Xmax, we study the same set of simulated showers used to derive
the primary energy function.  The fake data libraries use the four
primary types and
four hadronic interaction models processed through the BLANCA detector
simulation.  The simulated Cherenkov
lateral distributions are fit to the function in
Equation~\ref{eq.ldf}, which is exponential with a slope $s$ from
30\,m to 120\,m from the core.  The combined shower and
detector simulations show that this inner slope
is linearly related to \Xmax\ except for the deepest developing showers;
an additional small quadratic term is required for slopes exceeding
$s_\star=0.018\unit{m}^{-1}$.

% Xmax results plot
\begin{figure}
\bigfig{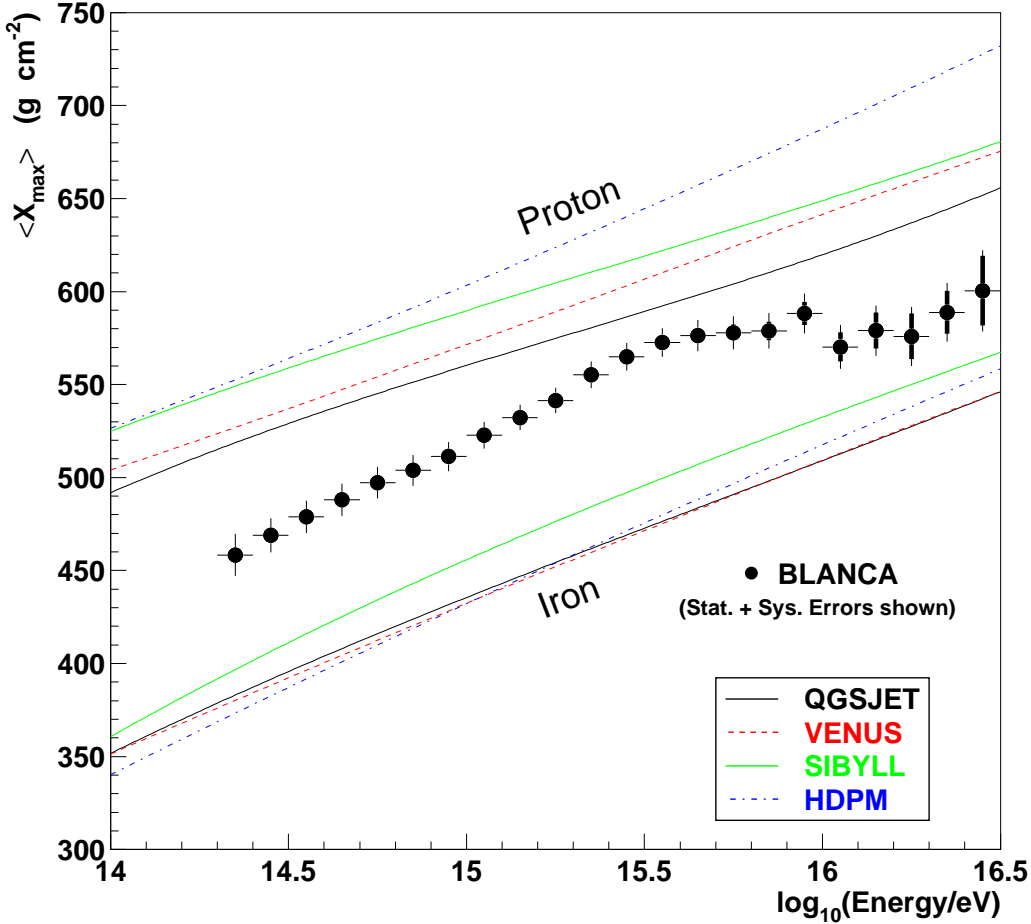}
\caption{The mean depth of shower maximum \meanXmax\ measured by CASA-BLANCA
as a function of energy.  The lines represent the values for pure 
proton or pure iron samples predicted by each of the four hadronic
interaction models.  The thick error bars represent the statistical
uncertainty, while the thin errors also include the 
systematic uncertainties discussed in the text; the former are
important only above 10\,PeV.}
\label{fig.xmax}
\end{figure}

%\P: Discussion of Xmax plot: data and predictions
The mean \Xmax\  is shown in Figure~\ref{fig.xmax}
as a function of energy. Both quantities are derived from the CASA-BLANCA data
using the CORSIKA/QGSJET Monte Carlo results.  We indeed find that the
results are very similar if any of the other hadronic models are used instead.
Numerical results are given in Table~\ref{tab.xmax}.
Figure~\ref{fig.xmax} also shows the mean \Xmax\  expected for pure samples of
proton primaries and iron primaries.  SIBYLL generally predicts deeper shower
maximum than other models, while HDPM exhibits a steeper elongation
rate than the others ($\sim\gcm{90}$ per decade compared with
70--$\gcm{75}$ typical of the other models).  The BLANCA results are
clearly consistent with a mixed composition throughout the energy
range, regardless of the preferred hadronic model.  The data suggest
that the composition becomes lighter approaching the knee and then
becomes heavier at higher energies.

\begin{figure}[t]
\bigfig{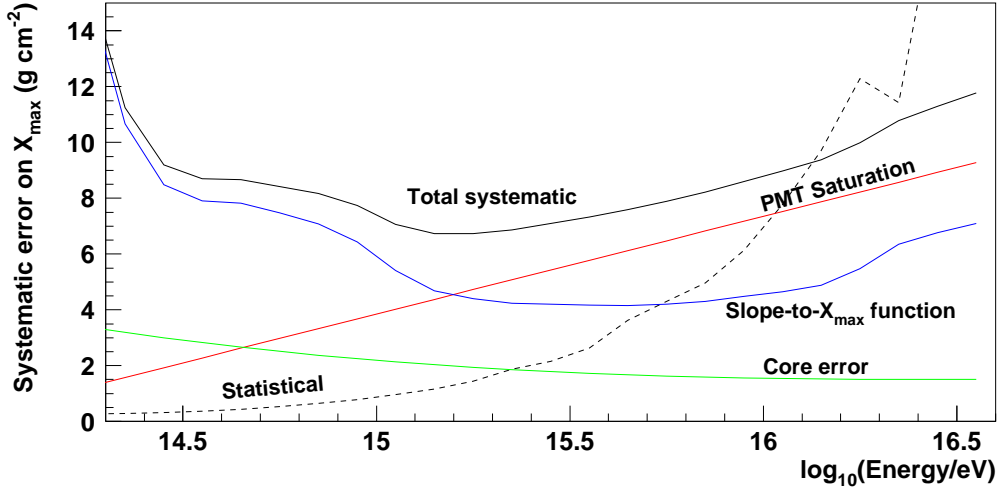}
\caption
{Systematic uncertainties on CASA-BLANCA \Xmax\  estimates.  At low
energies, the dominant error comes from the slope-to-\Xmax\  conversion
function;  at high energies, the 
uncertain photomultiplier linearity is more important.  For
comparison, the statistical error  on the CASA-BLANCA mean \Xmax\
measurements is also shown (dashed line), assuming a bin width of 0.1
decades.}
\label{fig.xmax_systematic}
\end{figure}

%\P: Systematics on Xmax
The shower depth estimated from Cherenkov observations is subject to a
number of small systematic uncertainties.
Random core errors lead to a 
systematic flattening of the Cherenkov inner slope, but
the effect is only a very small bias toward deeper \Xmax.
Photomultiplier saturation poses a potential problem at high
energies. However, the nonlinearity has been characterized in laboratory
studies of fourteen BLANCA detectors and the data corrected for its
effects.  The uncertainty on this correction leads to a systematic error in
\Xmax\  which is only \gcm{10} at the highest energies.  A third error
dominates at energies below 1\,PeV\@.  This error arises from the
limitations of the function which converts Cherenkov slope to an
estimate of \Xmax.  The function tends to overestimate the depth of
showers at the extreme ends of the BLANCA energy range.  It is difficult to
overcome this weakness without introducing at the same time a much
larger bias which depends on \Xmax\  itself.  Instead, we
take the error found in Monte Carlo studies as a
systematic error on the measured \Xmax.
The independent errors are added in quadrature to find the total
systematic error (Figure~\ref{fig.xmax_systematic}). 
Systematic errors are important only below 10\,PeV\@.  At high energy,
statistical errors are the more serious limitation on measuring mean
\Xmax.

%%%%%%%%%%%%%%%%%%%%%%%%%%%%%%%%%%%%%%%%%%%%%%%%%%%%%%%%%%%%%%%%%%%%%%%%%%%%%%%
\section{Mean Nuclear Mass} \label{sec.meanlna}

% Mean mass is closer to what it's really all about.
The mean depth of shower maximum results presented above
are essentially independent of a particular hadronic interaction model.
They are therefore useful for comparison with other experiments and
re-interpretation on the basis of future hadronic interaction models.
However, a quantity which is of much more direct astrophysical interest
is the mean nuclear mass of the primary cosmic rays.

%\P: Why we go ln(A)=f(s) and not ln(A)=f(Xmax)
We choose to derive mean primary mass directly from the Cherenkov
lateral distribution slope $s$.
It would make little difference if we were to do so via the
\Xmax\ values discussed in the previous section.
The important point is that while the transfer function from
$s$ to \Xmax\ is rather independent of the hadronic interaction
model, any interpretation in terms of absolute nuclear mass
is not.
This is clear from the disparity among the models of the mean proton
and iron \Xmax\  values shown in Figure~\ref{fig.xmax}.

% The ln(A) transfer function
At fixed primary energy, $s$ and \Xmax\
both depend linearly on the logarithm of nuclear mass
$A$, as do most composition-sensitive air shower parameters.
Following previous authors, we choose to work with the natural log, \lnA.
Unlike  $X_{max}$=$f_1(s)$, the transfer function $\ln(A)$=$f_2(s)$ depends
on energy.
Therefore we divide the Monte Carlo fake data into six bands of
$C_{120}$ and perform a linear fit to \lnA\ versus $s$ in each.
To interpret each real event the slope and intercept
of the transfer function are interpolated between the appropriate
bracketing bands.

% Errors on the transfer function
This method has little systematic bias apart from the differences between
hadronic interaction models.  The mean reconstructed \lnA\
for a pure sample of each simulated primary species is accurate to
to $\pm20\%$ over the BLANCA energy range.  On the other
hand, the random error is 
large on any single measurement of \lnA.  Shower-by-shower estimates of
primary mass cannot be made with any accuracy --- the fluctuations
inherent in the air shower process preclude them.  Nevertheless,
the mean value of \lnA\  provides a useful indicator of the cosmic ray
mass composition. 

% ln(A) results plot
\begin{figure}
\bigfig{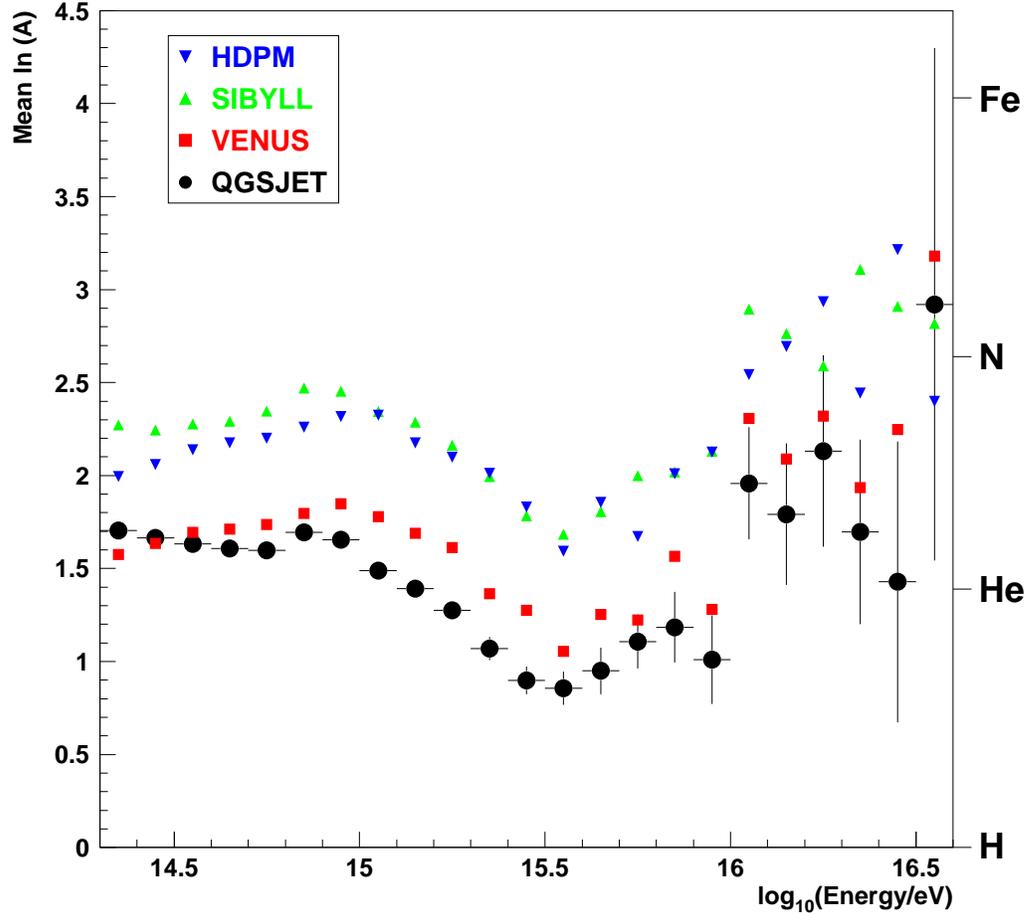}
\caption
{The mean logarithmic mass \meanlnA\ measured by CASA-BLANCA as a function
of energy.
The four sets of symbols show the BLANCA data interpreted using CORSIKA
coupled with the indicated hadronic interaction model.
Error bars are
statistical only and shown only on the QGSJET results; errors on the
other points are of similar size.  Assuming a hadronic model, the
systematic uncertainties in the \lnA\  estimate are typically 0.2.
The dominant systematic effect is clearly the difference
\emph{between} the hadronic models.}
\label{fig.lna}
\end{figure}

%\P: Discussion of ln(A) plot
The mean \lnA\  is shown as a function of primary energy
in Figure~\ref{fig.lna}.  The four sets of symbols show the BLANCA data
interpreted using CORSIKA coupled with each of the hadronic codes.
Numerical results are given in Table~\ref{tab.lna}.
The dependence on interaction model is
clear: the models set the overall mass scale differently, but they
indicate the same mass variation with energy.  The mean mass becomes
lighter with increasing energy through the knee, then becomes heavier
above $\sim 3\unit{PeV}$.  The \lnA\  plot exhibits the same trends
seen in the \Xmax\  results of the previous section.

%%%%%%%%%%%%%%%%% Multi-species fit
\section{A Multi-species Fit to the Cherenkov Data} \label{sec.multi}

%\P: multicomponent fit: the idea
The techniques described above involve estimating
\Xmax\  or \lnA\  for each shower and then taking the \emph{average}
over all showers in a given energy range.  By considering only the average
value we lose much of the available information.  Instead the
measured distribution of a composition sensitive parameter can be compared
with those predicted for a number of simulated primary species, providing a
more powerful technique to study cosmic ray composition.

% Binning data and pure species MC, smoothing
Comparing measured and simulated distributions requires high
statistics samples for both the real and Monte Carlo data sets.
We separate the data into five logarithmic energy bins between
$10^{14.5}$ and $10^{16.5}\unit{eV}$.  This bin choice is a compromise
between the need to include many showers in each range and the wish to
examine trends on as fine an energy scale as possible.
Within each range, we find the distribution of the
Cherenkov inner slope ($s$) for the real data and for pure samples
of each species in the Monte Carlo library (protons, He, N, and Fe).
The simulated distributions of $s$ (with detector effects) are
smoothed by a multiquadratic smoothing algorithm~\cite{hbook}.  To
preserve information about their limited statistics, the data
distributions are not smoothed.

% Fit: fractions not limited to 0-1 range and no constraint on sum
The multi-species fit in each energy range finds the
linear combination of the four simulated distributions which
reproduces the data distribution best.  Since each primary species has
a characteristic shape of its $s$ distribution, this fit uses
more information than simply the mean or even the width of $s$.  We do
not \emph{a priori} require the fractional contribution of each primary
type to lie in the physical range of 0--100\%, nor is the sum of the
fractions constrained to equal 1.0.  In practice, however, the sum
is always in the range $100\pm0.5\%$.
The fits use a MINUIT-based log-likelihood maximization
procedure~\cite{minuit}, which accounts properly for the Poisson
probability distribution of data events in bins with low statistics.

% Example multi comp fit
\begin{figure}
\bigfig{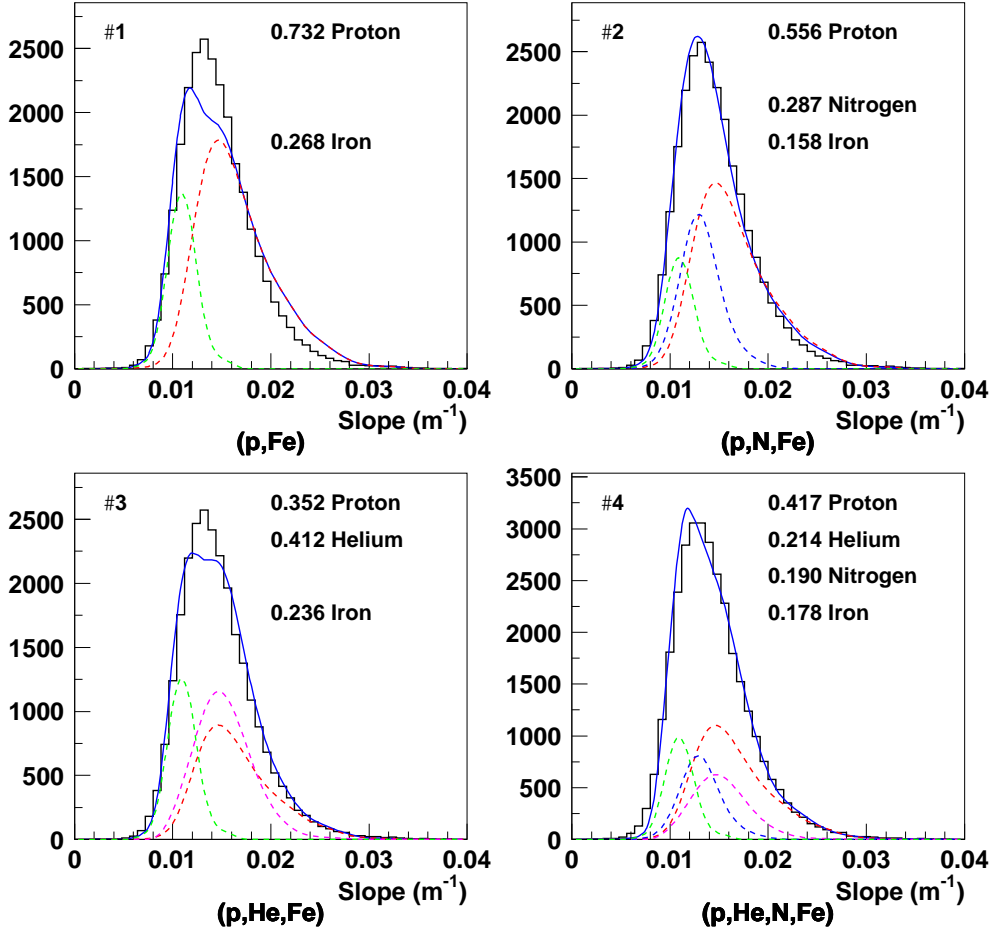}
\caption
{A demonstration of the multi-species fitting procedure, in this case the
energy range $10^{14.9}-10^{15.3}\unit{eV}$.  The solid histogram
gives the distribution of Cherenkov slope $s$ for all CASA-BLANCA data in
that energy range and is the same in all four panels.  The solid
curves show the combinations of Monte Carlo (QGSJET) showers which best
reproduce the measured distribution.  Panels 1--4 correspond to
different combinations of primary species.
Clearly a nitrogen component is needed to match the data, but a
helium component is also important.}
\label{fig.multi_example}
\end{figure}

%\P: requires Helium, Nitrogen
As an example, Figure~\ref{fig.multi_example} shows the multi-species
fit in the energy range $10^{14.9}$--$10^{15.3}\unit{eV}$.  The lower
right panel (\#4) displays the full fit using all four available
primary types. The other panels show the best fits that can be made
when helium (\#2), nitrogen (\#3), or both (\#1) are omitted.  The
Monte Carlo predictions cannot match the data using protons and iron
alone; the intermediate mass nitrogen species is also required.  Panel
\#2 shows that the best fit of p, N, and Fe to the data is very close but
fails to match the shape at the peak and in the long tail to high
values of $s$.  The data strongly suggest that at least the four
primary types considered here contribute to the cosmic rays just
below the knee.

% Multi comp results for QGSJET and VENUS
\begin{figure}[t]
\bigfig{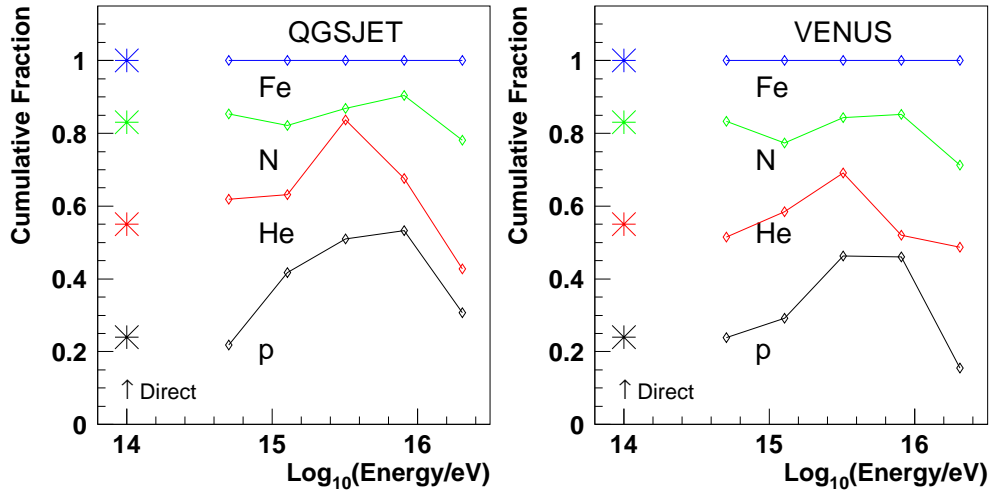}
\caption
{Results of the multi-species fit to the CASA-BLANCA data.  The line
graphs indicate the mixture of proton, helium, nitrogen, and iron
primaries which best reproduces the Cherenkov slope distributions of
the data.  The lines show \emph{cumulative} fractions (\emph{i.e.}\
the lowest line gives the proton fraction, while the next line gives
the combined proton and helium fraction).
QGSJET (\emph{left}) and VENUS (\emph{right}) are shown.  The other two
interaction models show similar trends but with heavier overall
composition.  The stars ($\ast$) at $10^{14}\unit{eV}$
show the results of direct measurements from the JACEE balloon-borne
emulsion experiment~\cite{jacee97,watson_rapp}. JACEE Ne-Si data have
been divided evenly into the N and Fe groups for comparison with BLANCA.}
\label{fig.multi_results}
\end{figure}

%\P: results of full 4-component
The fits demonstrated in Figure~\ref{fig.multi_example} (lower right)
are performed for all five energy ranges and using the predictions of
all four high energy hadronic interaction models.
Gauging the goodness of fit presents a problem.
As stated above the multi species fit uses a log-likelihood maximization
procedure; although this is an appropriate method for extracting abundance
fractions from the binned data, the actual value of the likelihood
is not useful~\cite{minuit}.
Therefore we calculate and use the familiar $\overline{\chi}^2$ quantity as
a rough guide to the fit quality.
In the lowest energy range,
the number of data events is so large that no combination of the four
cosmic ray species can reproduce the data adequately.
This is probably the result of limitations of the shower and detector
simulations, although it could also be due to the limited number of
species considered. Conversely, the
high energy ranges have too few events to constrain the abundances well.
The results for all models are presented in Table~\ref{tab.multi_species}.

% Discuss figure and composition implied
The results of the multi-species fit to the BLANCA Cherenkov slope
data are shown in Figure~\ref{fig.multi_results} for the 
QGSJET and VENUS models.  The SIBYLL and HDPM models show similar
trends but a heavier overall composition.  These latter two models
also give unphysical negative helium abundances in at least one energy
bin, and systematically poorer fits.
At 100\,TeV, data
from the JACEE balloon direct measurements~\cite{jacee97,watson_rapp}
are shown for comparison.  The direct composition
at 100\,TeV agrees well with  the BLANCA data at
400\,TeV\@.  
The results of the multi-species fit also agree with the
mean \Xmax\  and mean \lnA\  derived in the previous two sections.
All three ways of interpreting the data indicate that the cosmic ray
composition is lighter near 3\,PeV than it is at either 300\,TeV or 30\,PeV.

%%%%%%%%%%%%%%%%%%%%%%%%%%%%%%%%%%%%%%%%%%%%%%%%%%%%%%%%%%%%%%%%%%%%%%%%%%%%%%%
\section{Conclusions} \label{sec.conclusions}

% Restate the experiment, that we observe a knee
The CASA-BLANCA experiment has studied cosmic rays in the energy range
0.3--30\,PeV\@.  The primary energy and mass are found by
measuring the Cherenkov lateral distribution for each air shower.
In an effort to understand how results depend on the unknown physics
of high energy nuclear interactions, we have interpreted the data using
the CORSIKA air shower Monte Carlo program with four different hadronic
interaction models: QGSJET, VENUS, SIBYLL, and HDPM.

% Energy spectrum
The BLANCA energy spectrum agrees well with previous measurements and
exhibits a smooth knee near 2--3\,PeV in primary energy.
The model dependence of the energy scale is less than the absolute
calibration uncertainty.

\begin{figure}
\bigfig{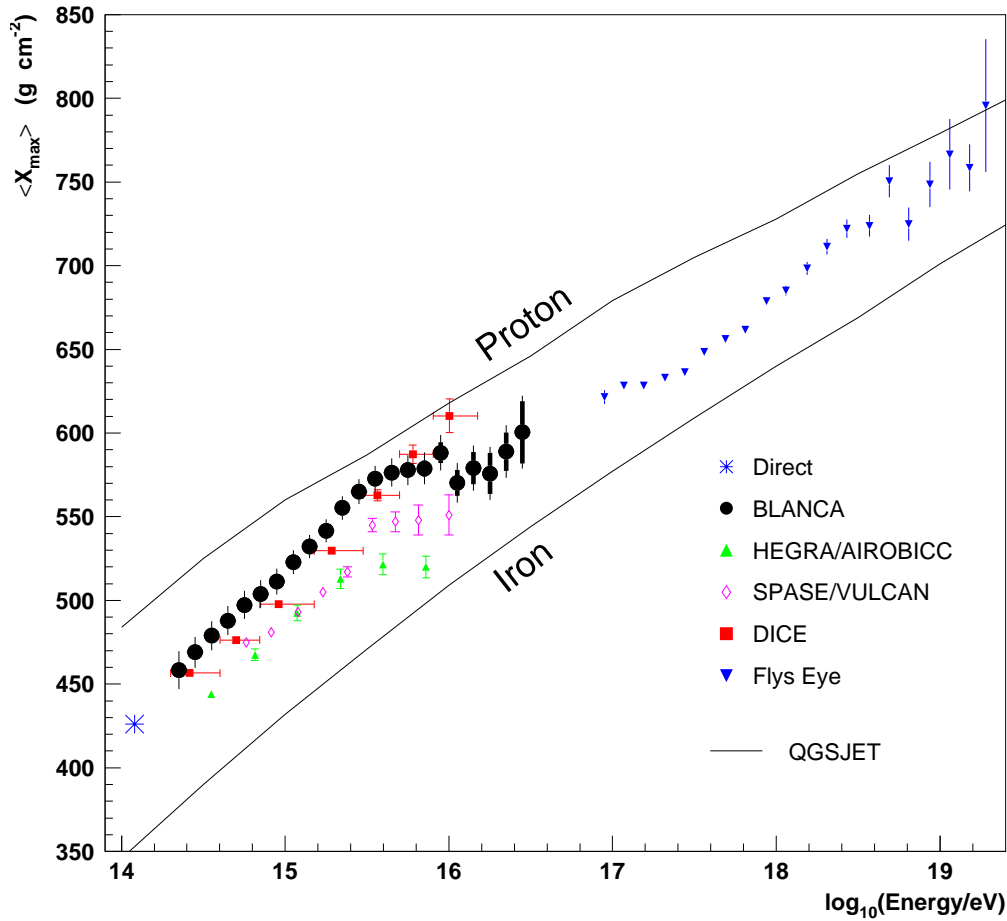}
\caption
{The CASA-BLANCA measurement of \Xmax\  compared with other results.  All
experiments operating near the knee use atmospheric Cherenkov
light, including DICE~\cite{dice}, the AIROBICC array of
HEGRA~\cite{airobicc}, and the VULCAN array at the South Pole~\cite{vulcan}.
The high energy measurements ($>10^{17}\unit{eV}$) by Fly's Eye use the
atmospheric fluorescence technique~\cite{flys_eye_result}.  The
``direct'' point estimates the mean \Xmax\  that would be expected on
the basis of direct balloon measurements~\cite{dice}.  The Monte
Carlo lines use CORSIKA with QGSJET~\cite{clem_corsika}.}
\label{fig.xmax_all}
\end{figure}

% Xmax model independent, and compare to other expt.
We find the transformation from measured Cherenkov lateral
distribution slope to the depth of shower maximum \Xmax\ to
be essentially model independent.
In Figure~\ref{fig.xmax_all} our results are compared to
previous experiments over a wide energy range.
The BLANCA data are well within the physically reasonable range bounded
by the pure proton and iron curves; furthermore they are consistent at
low energy with those expected from direct measurements and at high
energy with the Fly's Eye result~\cite{flys_eye_result}.

% mean ln(A)
We have also interpreted our data as a mean nuclear mass.
This is essentially equivalent to the \Xmax\ analysis but
is a quantity of more direct astrophysical interest.

% Choosing a model and getting the composition simultaneously
It has been a long held goal in the air shower field
to choose an adequate hadronic interaction model \emph{and} determine
the nuclear composition of the primary cosmic rays simultaneously.
With the advent of the powerful simulation tool provided by the CORSIKA group,
and high collecting power arrays such as CASA-BLANCA, it seems that this
ambition may be becoming a reality.
A multi component fit of the type described in Section~\ref{sec.multi}
is a much more efficient use of the available data than simply considering
the mean value and spread of a quantity, and the experimental statistics
are starting to justify this approach.
The agreement between data and simulation in Figure~\ref{fig.multi_example}
is impressive.

% What our result is and getting heavy beyond the knee
On the basis of our data we favor the QGSJET and VENUS models
and reject SIBYLL and HDPM.
At the same time, both of the ways in which we have analyzed our data
indicate that the cosmic ray composition is lighter near 3\,PeV than
it is at either 300\,TeV or 30\,PeV.
The trend towards heavier primary mass above 3\,PeV agrees with
the canonical model of Galactic production and a rigidity-dependent time
for escape, and is not consistent with acceleration at sites
such as AGN, which require a pure proton composition well above the
knee~\cite{agn_protheroe}.

% Getting lighter before the knee
The trend shown in our data to a lighter composition approaching the knee
is puzzling but not without theoretical precedent.
Swordy, arguing that there must be a minimum path length
in the Galaxy even for the highest energy cosmic rays, predicts a
light composition at the knee~\cite{swordy_model}.

%%%%%%%%%%%%%%%%%%%%%%%%%%%%%%%%%%%%%%%%%%%%%%%%%%%%%%%%%%%%%%%%%%%%%%%%%%%%%%%
% Acknowledgements
We acknowledge the invaluable assistance of the CASA-MIA
collaboration, as well as the University of Utah
High-Resolution Fly's Eye (HiRes) group and the command and staff of
the U.S.\ Army Dugway Proving Ground.  We thank
D.\ Heck and the rest of the CORSIKA team for providing and maintaining
their excellent program, and the authors of the hadronic interaction
models to which it is linked.  We thank C.\ Cassidy, J.\ Jacobs,
J.\ Meyer, M.\ Pritchard, and K.\ Riley for helping with BLANCA's
construction and K.\ Anderson and C.\ Eberhardy for calibration work.
We especially wish to thank M.\ Cassidy for his essential contributions
as our technician.  JF and CP acknowledge fellowships from the William
Grainger Foundation and the Robert R.\ McCormick Foundation,
respectively.  This work was supported by the U.S.\ National Science
Foundation. We would also like to thank S.\ Swordy for useful
conversations.

%%%%%%%%%%%%%%%%%%%%%%%%%%%%%%%%%%%%%%%%%%%%%%%%%%%%%%%%%%%%%%%%%%%%%%%%%%%%%%%

\raggedbottom
\pagebreak

%%%%%%%%%%%%%%%%%%%%%%%%%%%%%%%%%%%%%%%%%%%%%%%%%%%%%%%%%%%%%%%%%%%%%%%%%%%%%%%
\appendix
\section{Data tables}

% The table of raw spectrum data
\begin{table}[b]
\begin{center}
\begin{tabular}{cc}
Energy range & Differential flux $J(E)$ in each bin \\
$\log_{10}(E/1\unit{eV})$ &
{(m$^{-2}$\,sr$^{-1}$\,s$^{-1}$\,GeV$^{-1}$)} \\  \hline
 % This table auto-generated on cygnus.uchicago.edu by final_plots.kumac 
 % on 23/12/99 at 08.49.52 UT.
%
% Modified by Joe on 6 March 2000 to remove highest bin, 
%
 % 
 14.3 -- 14.4  & $ (12.12 \pm 0.04) \times 10^{-11}$ \\ 
 14.4 -- 14.5  & $ (6.68 \pm 0.02) \times 10^{-11}$ \\ 
 14.5 -- 14.6  & $ (3.58 \pm 0.02) \times 10^{-11}$ \\ 
 14.6 -- 14.7  & $ (1.91 \pm 0.01) \times 10^{-11}$ \\ 
 14.7 -- 14.8  & $ (1.03 \pm 0.01) \times 10^{-11}$ \\ 
 14.8 -- 14.9  & $ (5.42 \pm 0.04) \times 10^{-12}$ \\ 
 14.9 -- 15.0  & $ (2.90 \pm 0.03) \times 10^{-12}$ \\ 
 15.0 -- 15.1  & $ (1.57 \pm 0.02) \times 10^{-12}$ \\ 
 15.1 -- 15.2  & $ (8.18 \pm 0.12) \times 10^{-13}$ \\ 
 15.2 -- 15.3  & $ (4.36 \pm 0.08) \times 10^{-13}$ \\ 
 15.3 -- 15.4  & $ (2.21 \pm 0.05) \times 10^{-13}$ \\ 
 15.4 -- 15.5  & $ (1.22 \pm 0.03) \times 10^{-13}$ \\ 
 15.5 -- 15.6  & $ (6.2 \pm 0.2) \times 10^{-14}$ \\ 
 15.6 -- 15.7  & $ (2.9 \pm 0.1) \times 10^{-14}$ \\ 
 15.7 -- 15.8  & $ (1.5 \pm 0.1) \times 10^{-14}$ \\ 
 15.8 -- 15.9  & $ (7.7 \pm 0.5) \times 10^{-15}$ \\ 
 15.9 -- 16.1  & $ (2.9 \pm 0.2) \times 10^{-15}$ \\ 
 16.1 -- 16.3  & $ (8.1 \pm 0.8) \times 10^{-16}$ \\ 
 16.3 -- 16.5  & $ (2.1 \pm 0.3) \times 10^{-16}$ \\ 
 16.5 -- 16.7  & $ (3.1 \pm 1.0) \times 10^{-17}$ \\ 
 16.7 -- 16.9  & $ (2.3 \pm 0.7) \times 10^{-17}$ \\ 
%
% Old version:
% 
%  14.3 -- 14.4  & $ 1.21 \times 10^{-10} \pm 3.60 \times 10^{-13}$ \\ 
%  14.4 -- 14.5  & $ 6.68 \times 10^{-11} \pm 2.39 \times 10^{-13}$ \\ 
%  %  
%  14.5 -- 14.6  & $ 3.58 \times 10^{-11} \pm 1.56 \times 10^{-13}$ \\ 
%  14.6 -- 14.7  & $ 1.91 \times 10^{-11} \pm 1.01 \times 10^{-13}$ \\ 
%  14.7 -- 14.8  & $ 1.03 \times 10^{-11} \pm 6.64 \times 10^{-14}$ \\ 
%  14.8 -- 14.9  & $ 5.42 \times 10^{-12} \pm 4.29 \times 10^{-14}$ \\ 
%  14.9 -- 15.0  & $ 2.90 \times 10^{-12} \pm 2.80 \times 10^{-14}$ \\ 
%  15.0 -- 15.1  & $ 1.57 \times 10^{-12} \pm 1.83 \times 10^{-14}$ \\ 
%  15.1 -- 15.2  & $ 8.18 \times 10^{-13} \pm 1.18 \times 10^{-14}$ \\ 
%  15.2 -- 15.3  & $ 4.36 \times 10^{-13} \pm 7.69 \times 10^{-15}$ \\ 
%  15.3 -- 15.4  & $ 2.21 \times 10^{-13} \pm 4.88 \times 10^{-15}$ \\ 
%  15.4 -- 15.5  & $ 1.22 \times 10^{-13} \pm 3.23 \times 10^{-15}$ \\ 
%  %  
%  15.5 -- 15.6  & $ 6.19 \times 10^{-14} \pm 2.05 \times 10^{-15}$ \\ 
%  15.6 -- 15.7  & $ 2.86 \times 10^{-14} \pm 1.24 \times 10^{-15}$ \\ 
%  15.7 -- 15.8  & $ 1.51 \times 10^{-14} \pm 8.02 \times 10^{-16}$ \\ 
%  15.8 -- 15.9  & $ 7.68 \times 10^{-15} \pm 5.10 \times 10^{-16}$ \\ 
%  15.9 -- 16.1  & $ 2.95 \times 10^{-15} \pm 1.92 \times 10^{-16}$ \\ 
%  16.1 -- 16.3  & $ 8.14 \times 10^{-16} \pm 8.08 \times 10^{-17}$ \\ 
%  16.3 -- 16.5  & $ 2.12 \times 10^{-16} \pm 3.26 \times 10^{-17}$ \\ 
%  16.5 -- 16.7  & $ 3.12 \times 10^{-17} \pm 9.97 \times 10^{-18}$ \\ 
%  16.7 -- 16.9  & $ 2.31 \times 10^{-17} \pm 6.97 \times 10^{-18}$ \\ 
%  % 16.9 -- 17.1  & $ 2.31 \times 10^{-18} \pm 1.64 \times 10^{-18}$ \\ 

\end{tabular}
\end{center}
\caption
{The primary cosmic ray energy spectrum measured by CASA-BLANCA.
Bin widths rise with increasing energy so that $E_{max}/E_{min}=
10^{0.1}$ at lower energies, while $E_{max}/E_{min}=10^{0.2}$ for the
five highest bins.  Errors represent only the Poisson
uncertainty in each bin.  There is an additional instrumental systematic
uncertainty of 18\%.
These results use the QGSJET-derived energy transfer function.}
\label{tab.spectrum}
\end{table}

% Table of mean s and Xmax
\begin{table}[p]
\begin{center}
\begin{tabular}{cll}
Energy range & Mean $s$ & $ \meanXmax\pm\unit{stat.}\pm\unit{sys.}$\\
$\log_{10}(E/\unit{eV})$ & 
$(10^{-3}\unit{m}^{-1})$ \hspace{1.3cm} & 
$(\gcm{})$ \\  \hline
 % This table auto-generated on cygnus.uchicago.edu by final_plots.kumac 
 % on 24/11/99 at 04.15.38 UT.
 % 
 % Modified 6 March 2000 by Joe to get significant digits reasonable.
 %
 14.3 -- 14.4  & $ 11.4 \pm 0.0$ & $ 458 \pm 0.3 \pm 12$ \\ 
 14.4 -- 14.5  & $ 11.9 \pm 0.0$ & $ 469 \pm 0.3 \pm 10$ \\ 
 14.5 -- 14.6  & $ 12.5 \pm 0.0$ & $ 479 \pm 0.4 \pm  9$ \\ 
 14.6 -- 14.7  & $ 12.9 \pm 0.0$ & $ 488 \pm 0.4 \pm  9$ \\ 
 14.7 -- 14.8  & $ 13.4 \pm 0.0$ & $ 497 \pm 0.5 \pm  9$ \\ 
 14.8 -- 14.9  & $ 13.7 \pm 0.0$ & $ 504 \pm 0.6 \pm  9$ \\ 
 14.9 -- 15.0  & $ 14.1 \pm 0.0$ & $ 511 \pm 0.8 \pm  8$ \\ 
 15.0 -- 15.1  & $ 14.7 \pm 0.0$ & $ 523 \pm 1 \pm 7$ \\ 
 15.1 -- 15.2  & $ 15.2 \pm 0.1$ & $ 532 \pm 1 \pm 7$ \\ 
 15.2 -- 15.3  & $ 15.6 \pm 0.1$ & $ 542 \pm 1 \pm 7$ \\ 
 15.3 -- 15.4  & $ 16.3 \pm 0.1$ & $ 555 \pm 2 \pm 7$ \\ 
 15.4 -- 15.5  & $ 16.8 \pm 0.1$ & $ 565 \pm 2 \pm 7$ \\ 
 15.5 -- 15.6  & $ 17.2 \pm 0.1$ & $ 573 \pm 3 \pm 7$ \\ 
 15.6 -- 15.7  & $ 17.3 \pm 0.2$ & $ 576 \pm 4 \pm 8$ \\ 
 15.7 -- 15.8  & $ 17.4 \pm 0.2$ & $ 578 \pm 4 \pm 8$ \\ 
 15.8 -- 15.9  & $ 17.5 \pm 0.2$ & $ 579 \pm 5 \pm 8$ \\ 
 15.9 -- 16.0  & $ 17.9 \pm 0.3$ & $ 588 \pm 6 \pm 9$ \\ 
 16.0 -- 16.1  & $ 17.0 \pm 0.4$ & $ 570 \pm  8 \pm  9$ \\ 
 16.1 -- 16.2  & $ 17.5 \pm 0.4$ & $ 579 \pm 10 \pm  9$ \\ 
 16.2 -- 16.3  & $ 17.4 \pm 0.6$ & $ 576 \pm 12 \pm 10$ \\ 
 16.3 -- 16.4  & $ 18.1 \pm 0.5$ & $ 589 \pm 11 \pm 11$ \\ 
 16.4 -- 16.5  & $ 18.5 \pm 0.8$ & $ 600 \pm 19 \pm 11$ \\ 
 16.5 -- 16.6  & $ 17.2 \pm 1.5$ & $ 570 \pm 31 \pm 12$ \\

\end{tabular}
\end{center}
\caption
{The mean Cherenkov inner slope ($s$) measured by CASA-BLANCA and the
corresponding mean \Xmax.
The first column of errors given for each quantity is statistical;
the standard deviation divided by $\sqrt{N}$.
For \Xmax\ a systematic error is given which is due to a combination
of effects (see Section~\ref{sec.xmax}).
These results use the QGSJET-derived \Xmax\ transfer function.}
\label{tab.xmax}
\end{table}

% Table of mean ln(A)
\begin{table}[p]
\begin{center}
\begin{tabular}{cllll}
Energy range & \multicolumn{4}{c}{Mean \lnA} \\
$\log_{10}(E/1\unit{eV})$ & QGSJET & VENUS & SIBYLL & HDPM \\  \hline
 % This table auto-generated on cygnus.uchicago.edu by final_plots.kumac 
 % on 24/11/99 at 04.16.43 UT.
 % 
 % Modified 6 March 2000 by Joe to get significant digits reasonable.
 %
 14.3 -- 14.4  & $ 1.71 \pm 0.01$ & 1.58 & 2.27 & 1.99   \\ 
 14.4 -- 14.5  & $ 1.67 \pm 0.01$ & 1.63 & 2.24 & 2.06   \\ 
 14.5 -- 14.6  & $ 1.63 \pm 0.01$ & 1.69 & 2.28 & 2.14   \\ 
 14.6 -- 14.7  & $ 1.61 \pm 0.01$ & 1.71 & 2.29 & 2.18   \\ 
 14.7 -- 14.8  & $ 1.60 \pm 0.02$ & 1.74 & 2.35 & 2.20   \\ 
 14.8 -- 14.9  & $ 1.69 \pm 0.02$ & 1.80 & 2.47 & 2.26   \\ 
 14.9 -- 15.0  & $ 1.65 \pm 0.03$ & 1.85 & 2.45 & 2.32   \\ 
 15.0 -- 15.1  & $ 1.49 \pm 0.03$ & 1.78 & 2.34 & 2.32   \\ 
 15.1 -- 15.2  & $ 1.39 \pm 0.04$ & 1.69 & 2.29 & 2.18   \\ 
 15.2 -- 15.3  & $ 1.28 \pm 0.05$ & 1.61 & 2.16 & 2.10   \\ 
 15.3 -- 15.4  & $ 1.07 \pm 0.06$ & 1.36 & 1.99 & 2.01   \\ 
 15.4 -- 15.5  & $ 0.90 \pm 0.07$ & 1.28 & 1.78 & 1.83   \\ 
 15.5 -- 15.6  & $ 0.86 \pm 0.09$ & 1.06 & 1.68 & 1.59   \\ 
 15.6 -- 15.7  & $ 0.95 \pm 0.12$ & 1.25 & 1.81 & 1.85   \\ 
 15.7 -- 15.8  & $ 1.1 \pm 0.1$ & 1.2 & 2.0 & 1.7   \\ 
 15.8 -- 15.9  & $ 1.2 \pm 0.2$ & 1.6 & 2.0 & 2.0   \\ 
 15.9 -- 16.0  & $ 1.0 \pm 0.2$ & 1.3 & 2.1 & 2.1   \\ 
 16.0 -- 16.1  & $ 2.0 \pm 0.3$ & 2.3 & 2.9 & 2.5   \\ 
 16.1 -- 16.2  & $ 1.8 \pm 0.4$ & 2.1 & 2.8 & 2.7   \\ 
 16.2 -- 16.3  & $ 2.1 \pm 0.5$ & 2.3 & 2.6 & 2.9   \\ 
 16.3 -- 16.4  & $ 1.7 \pm 0.5$ & 1.9 & 3.1 & 2.4   \\ 
 16.4 -- 16.5  & $ 1.4 \pm 0.8$ & 2.2 & 2.9 & 3.2   \\ 
 16.5 -- 16.6  & $ 2.9 \pm 1.4$ & 3.2 & 2.8 & 2.4   \\ 

\end{tabular}
\end{center}
\caption{The mean \lnA\  measured by CASA-BLANCA.
The four columns show the data interpreted according to
each hadronic interaction model.
The variation provides some insight into the systematic errors.
Statistical errors shown on the QGSJET points are similar for all four models.} 
\label{tab.lna}
\end{table}

% Table of multi component values
\begin{table}[p]
\begin{center}
\begin{tabular}{cccccr}
Energy Range & \multicolumn{4}{c}{\dotfill Abundance (\%)\dotfill} & 
$\overline{\chi}^2$ of \\
$\log_{10}(E/1\unit{eV})$ & p & He & N & Fe & Fit \\ \hline
\multicolumn{6}{c}{QGSJET} \\
 % This table auto-generated on cygnus.uchicago.edu by final_comp.kumac 
 % on 30/11/99 at 05.41.02 UT.
 % 
 % Joe modified 6 March 2000 to remove extraneous insignificant digits.

 % QGSJET \\ 
 14.5 -- 14.9  & $ 21.8 \pm  0.4 $ & $ 40.1 \pm  0.7 $ & $ 23.4 \pm  0.7 $ & $ 14.6 \pm  0.3 $ &  38.1 \\ 
 14.9 -- 15.3  & $ 42 \pm   1 $ & $ 21 \pm   2 $ & $ 19 \pm   1 $ & $ 18 \pm  1 $ &   4.5 \\ 
 15.3 -- 15.7  & $ 51 \pm   3 $ & $ 33 \pm   4 $ & $  3 \pm   3 $ & $ 13 \pm  1 $ &   1.9 \\ 
 15.7 -- 16.1  & $ 53 \pm   8 $ & $ 14 \pm  10 $ & $ 23 \pm   6 $ & $ 10 \pm  3 $ &   0.7 \\ 
 16.1 -- 16.5  & $ 31 \pm  12 $ & $ 12 \pm  18 $ & $ 35 \pm  17 $ & $ 22 \pm  8 $ &   1.9 \\ \hline
% 14.9 -- 15.3  & $ 41.7 \pm  1.3 $ & $ 21.4 \pm  1.8 $ & $ 19.0 \pm  1.3 $ & $ 17.8 \pm  0.6 $ &   4.5 \\ 
% 15.3 -- 15.7  & $ 51.0 \pm  2.7 $ & $ 32.7 \pm  4.0 $ & $  3.2 \pm  2.9 $ & $ 13.1 \pm  1.1 $ &   1.9 \\ 
% 15.7 -- 16.1  & $ 53.3 \pm  7.6 $ & $ 14.3 \pm  9.5 $ & $ 22.8 \pm  6.4 $ & $  9.6 \pm  3.0 $ &   0.7 \\ 
% 16.1 -- 16.5  & $ 30.7 \pm  12. $ & $ 12.1 \pm  18. $ & $ 35.4 \pm  17. $ & $ 21.9 \pm  8.2 $ &   1.9 \\ \hline
 %  
 \multicolumn{6}{c}{VENUS} \\ 
 14.5 -- 14.9  & $ 23.9 \pm  0.4 $ & $ 27.6 \pm  0.7 $ & $ 31.8 \pm  0.5 $ & $ 16.7 \pm  0.3 $ &  47.8 \\ 
 14.9 -- 15.3  & $ 29 \pm  1 $ & $ 29 \pm   2 $ & $ 19 \pm  1 $ & $ 23 \pm  1 $ &   5.9 \\ 
 15.3 -- 15.7  & $ 46 \pm  2 $ & $ 23 \pm   4 $ & $ 15 \pm  3 $ & $ 16 \pm  1 $ &   1.7 \\ 
 15.7 -- 16.1  & $ 46 \pm  6 $ & $  6 \pm   9 $ & $ 33 \pm  7 $ & $ 15 \pm  3 $ &   0.8 \\ 
 16.1 -- 16.5  & $ 16 \pm  9 $ & $ 33 \pm  14 $ & $ 23 \pm 13 $ & $ 29 \pm  8 $ &   1.8 \\ \hline
%  14.9 -- 15.3  & $ 29.1 \pm  1.1 $ & $ 29.3 \pm  1.9 $ & $ 18.9 \pm  1.4 $ & $ 22.6 \pm  0.6 $ &   5.9 \\ 
%  15.3 -- 15.7  & $ 46.3 \pm  2.4 $ & $ 22.8 \pm  3.8 $ & $ 15.2 \pm  2.7 $ & $ 15.7 \pm  1.1 $ &   1.7 \\ 
%  15.7 -- 16.1  & $ 46.0 \pm  5.8 $ & $  5.9 \pm  9.4 $ & $ 33.3 \pm  7.0 $ & $ 14.8 \pm  2.6 $ &   0.8 \\ 
%  16.1 -- 16.5  & $ 15.5 \pm  8.9 $ & $ 33.1 \pm  14. $ & $ 22.6 \pm  13. $ & $ 28.8 \pm  7.8 $ &   1.8 \\
%
 \multicolumn{6}{c}{SIBYLL} \\ 
 14.5 -- 14.9  & $ 16.8 \pm  0.4 $ & $ 20.9 \pm  0.7 $ & $ 24.8 \pm  0.7 $ & $ 37.4 \pm  0.3 $ &  49.4 \\ 
 14.9 -- 15.3  & $ 35 \pm  1 $ & $ -7 \pm  2 $ & $ 39 \pm  1 $ & $ 33 \pm  1 $ &  21.6 \\ 
 15.3 -- 15.7  & $ 37 \pm  3 $ & $ 21 \pm  5 $ & $  9 \pm  4 $ & $ 33 \pm  2 $ &   4.8 \\ 
 15.7 -- 16.1  & $ 31 \pm  6 $ & $ 19 \pm 10 $ & $  9 \pm  8 $ & $ 41 \pm  4 $ &   1.2 \\ 
 16.1 -- 16.5  & $  9 \pm  9 $ & $ 31 \pm 17 $ & $  8 \pm 19 $ & $ 53 \pm 11 $ &   2.6 \\ \hline
%  14.9 -- 15.3  & $ 35.0 \pm  1.0 $ & $ -6.9 \pm  1.7 $ & $ 38.9 \pm  1.4 $ & $ 33.0 \pm  0.7 $ &  21.6 \\ 
%  15.3 -- 15.7  & $ 36.7 \pm  2.9 $ & $ 21.3 \pm  4.8 $ & $  9.2 \pm  3.6 $ & $ 32.7 \pm  1.5 $ &   4.8 \\ 
%  15.7 -- 16.1  & $ 31.3 \pm  5.5 $ & $ 19.4 \pm  9.5 $ & $  8.6 \pm  8.0 $ & $ 40.7 \pm  3.8 $ &   1.2 \\ 
%  16.1 -- 16.5  & $  8.6 \pm  9.0 $ & $ 31.4 \pm  17. $ & $  7.5 \pm  19. $ & $ 52.5 \pm 11. $ &   2.6 \\ \hline
 %  
 \multicolumn{6}{c}{HDPM} \\ 
 14.5 -- 14.9  & $ 19.9 \pm  0.3 $ & $ 17.9 \pm  0.5 $ & $ 31.4 \pm  0.5 $ & $ 30.7 \pm  0.2 $ &  90.0 \\ 
 14.9 -- 15.3  & $ 19 \pm  1 $ & $ 23 \pm  1 $ & $ 24 \pm  1 $ & $ 34 \pm  1 $ &  11.5 \\ 
 15.3 -- 15.7  & $ 32 \pm  2 $ & $ 16 \pm  3 $ & $ 26 \pm  2 $ & $ 26 \pm  1 $ &   3.2 \\ 
 15.7 -- 16.1  & $ 37 \pm  4 $ & $ -3 \pm  6 $ & $ 43 \pm  6 $ & $ 23 \pm  3 $ &   1.2 \\ 
 16.1 -- 16.5  & $ 12 \pm  6 $ & $ 21 \pm 12 $ & $ 18 \pm 12 $ & $ 49 \pm  8 $ &   1.5 \\ \hline
%  14.9 -- 15.3  & $ 18.6 \pm  0.6 $ & $ 23.2 \pm  1.1 $ & $ 24.3 \pm  0.9 $ & $ 34.0 \pm  0.5 $ &  11.5 \\ 
%  15.3 -- 15.7  & $ 31.7 \pm  1.5 $ & $ 16.3 \pm  2.5 $ & $ 26.4 \pm  2.1 $ & $ 25.6 \pm  1.0 $ &   3.2 \\ 
%  15.7 -- 16.1  & $ 36.7 \pm  3.9 $ & $ -3.0 \pm  6.4 $ & $ 43.4 \pm  5.5 $ & $ 22.9 \pm  2.5 $ &   1.2 \\ 
%  16.1 -- 16.5  & $ 11.9 \pm  5.8 $ & $ 21.2 \pm  12. $ & $ 18.0 \pm  12. $ & $ 48.9 \pm  7.7 $ &   1.5 \\ \hline %  

\end{tabular}
\end{center}
\caption{Results of the multi-species fits to the CASA-BLANCA data.
Statistical errors on each fraction are strongly correlated.  The errors
increase with energy due to limited statistics.
Unphysical negative abundances are a result of a poor hadronic model and/or
inadequate statistics.} 
\label{tab.multi_species}
\end{table}


\begin{thebibliography}{99}

\bibitem{Cesarsky} P. O. Lagage and C. J. Cesarsky, Astron.\ Astrophys.,
{\bf 118}, 223 (1983) and {\bf 125}, 249 (1983).

\bibitem{Drury94} L. O'C. Drury  \etal, {\em Astron.\ Astrophys.},
{\bf 287} (1994) 959.

\bibitem{Brennan58} M. H. Brennan  \etal, {\em Nature}, {\bf 182} (1958) 973.

\bibitem{Chudakov60} A. E. Chudakov \etal, {\em Proc.\ 6th Int.\ Conf.\
on Cosmic Rays, Moscow}, {\bf 2} (1960) 47.

%\bibitem{Gaisser94}
%T. K. Gaisser \etal,
%{\em Proc. 1994 Snowmass Summer Study}, 
%E. W. Kolb, R. D. Peccei, ed., World Scientific, Singapore, 1995.

\bibitem{patterson_hillas} J. R. Patterson and A. M. Hillas, J. Phys.\
G, {\bf 9}, 1433 (1983).

\bibitem{Dawson89} B. R. Dawson \etal, {\em J.\ Phys.\ G: Nucl.\ Part.\ Phys.},
{\bf 15} (1989) 893.

\bibitem{airobicc} F. Arqueros \etal, accepted in Astron.\ Astrophys.\
(1999).  Preprint astro-ph/9908202.

\bibitem{vulcan} J. E. Dickinson \etal, \emph{Proc.\ 26th Int.\ Cosmic
Ray Conf., Salt Lake City}, {\bf 3}, 136 (1999).

\bibitem{casa_nim} A. Borione \etal, {Nucl.\ Instrum.\ Meth.}, {\bf
A346}, 329 (1994).

\bibitem{blanca_instrument2} M. Cassidy \etal, \emph{Proc.\ 25th Int.\
Cosmic Ray Conf., Durban}, {\bf 5} , 189 (1997).

\bibitem{winston} W. T. Welford and R. Winston, \emph{High Collection
Non-imaging Optics}, Academic Press, San Diego, 1989.

\bibitem{thesis} J. W. Fowler, PhD Thesis, University of Chicago, 2000.

\bibitem{corsika} D. Heck \etal,  Report FZKA 6019,
Institut f\"ur Kern\-physik, For\-schungs\-zen\-trum Karls\-ruhe,
1998.

\bibitem{qgsjet} N. N. Kalmykov, S. S. Ostapchenko, and A. I. Pavlov,
Bull.\ Russ.\ Acad.\ Sci.\ (Physics), {\bf 58}, 1966 (1994).

\bibitem{venus} K. Werner, Phys.\ Reports, {\bf 232}, 87 (1993).

\bibitem{sibyll} R. S. Fletcher \etal, Phys.\ Rev.\ D, {\bf 50},
5710 (1994).

\bibitem{icrc_corsika} L. F. Fortson \etal, \emph{Proc.\ 26th Int.\
Cosmic Ray Conf., Salt Lake City}, {\bf 5}, 336 (1999).

\bibitem{dice} S. P. Swordy and D. B. Kieda, preprint
astro-ph/9909381, accepted in \emph{Astroparticle Phys.}, 1999.

\bibitem{ande_spec} M. A. K. Glasmacher, \etal, Astroparticle Phys.,
{\bf 10}, 291 (1999).

\bibitem{akeno} M. Nagano \etal, J. Phys.\ G, {\bf 10}, 1295 (1984).

\bibitem{tibet} M. Amenomori \etal, Astrophys.\ J., {\bf 461}, 408 (1996).

%\bibitem{knee_discovery} G. V. Kulikov and G. B. Khristiansen, 
% J.\ Exptl.\ Theoret.\ Phys.\ (USSR), {\bf 35}, 635 (1958).
% Appears in English translation as 
%Soviet Phys.\ JETP, {\bf 8}, 441 (1959).

\bibitem{flys_eye} R. M. Baltrusaitis \etal, Nucl.\ Instr.\ and
Meth.\ A {\bf 240}, 410 (1985).

\bibitem{PattersonHillas}J. R. Patterson and A. M. Hillas, J. Phys.\
G, {\bf 9}, 1433 (1983).

%\bibitem{ande_comp} M. A. K. Glasmacher \etal, {Astropart.\ Phys.}, 
%{\bf 12}, 1 (1999).

\bibitem{hbook} F. James, ``HBOOK Reference Manual, Version
4.24,'' CERN Program Library Long Writeup Y250, Geneva, 1995.

\bibitem{minuit} F. James, ``MINUIT Reference Manual,'' CERN Program
Library Long Writeup D506, Geneva, 1994.

\bibitem{jacee97} M. L. Cherry \etal, \emph{Proc.\ 25th Int.\ Cosmic
Ray Conf., Durban}, {\bf 4}, 1 (1997).

\bibitem{watson_rapp} A. A. Watson, \emph{Proc.\ 25th Int.\ Cosmic
Ray Conf., Durban}, {\bf 8}, 257 (1997).

\bibitem{flys_eye_result} D. Bird \etal, Phys.\ Rev.\ Lett., {\bf 71},
3401 (1993).

\bibitem{clem_corsika} C. Pryke, ``Shower Model Comparison I:
Longitudinal Profile,'' Pierre Auger project technical note 
GAP-98-035, FNAL, 1998.

\bibitem{agn_protheroe} R. J. Protheroe and A. P. Szabo, Phys.\ Rev.\
Lett., {\bf 69}, 2885 (1992).

\bibitem{swordy_model} S. Swordy, \emph{Proc.\ 24th Int.\ Cosmic Ray Conf.,
Rome}, {\bf 2}, 697 (1995).

\end{thebibliography}
\end{document}